\documentclass[journal]{IEEEtran}

\ifCLASSINFOpdf
\else
   \usepackage[dvips]{graphicx}
\fi
\usepackage{url}

\usepackage{cite}
\usepackage{amsmath,amssymb,amsfonts}

\usepackage{algorithm}
\usepackage{algorithmic}

\usepackage[dvipdfmx]{graphicx}

\usepackage{textcomp}

\usepackage{xcolor}
\usepackage{bm}
\usepackage{mathtools}
\usepackage{multirow}

\newcommand{\argmin}{\arg\,\min}

\begin{document}

\title{Approximate Proximal Operators for Analog Compressed Sensing Using PN-junction Diode }

\author{
    Soma Furusawa, 
    Taisei Kato, 
    Ryo Hayakawa, \IEEEmembership{Member, IEEE},
    and Kazunori Hayashi, \IEEEmembership{Member, IEEE}
    
\thanks{This work has been supported in part by JST CREST JPMJCR21C3 and 
JSPS KAKENHI Grant Number 25H01111.

Soma Furusawa and Kazunori Hayashi are with the Graduate School of Informatics, Kyoto University, Kyoto, Japan
(e-mail: furusawa.soma.82h@st.kyoto-u.ac.jp; hayashi.kazunori.4w@kyoto-u.ac.jp).

Taisei Kato is with the Graduate School of Engineering Science, Osaka University, Osaka, Japan 
(e-mail: taisei\_kato@sip.sys.es.osaka-u.ac.jp).

Ryo Hayakawa is with the Institute of Engineering, Tokyo University of Agriculture and Technology, Tokyo, Japan
(e-mail: hayakawa@go.tuat.ac.jp).}
}

\markboth{Journal of \LaTeX\ Class Files, Vol. 14, No. 8, AUGUST 2015}
{Shell \MakeLowercase{\textit{et al.}}: Bare Demo of IEEEtran.cls for IEEE Journals}

\maketitle

\begin{abstract}
In order to realize analog compressed sensing, the paper considers approximate proximal operators of the $\ell_1$ and minimax concave penalty (MCP) regularization functions. 
Specifically, we propose to realize the approximate functions by an electric analog circuit using forward voltage-current (V-I) characteristics of the PN-junction diodes.
To confirm the validity of the proposed approach, we employ the proposed approximate proximal operators for the $\ell_1$ and MCP regularization functions in compressed sensing with the proximal gradient method. 
The sparse reconstruction performance of the algorithms using the proposed approximate proximal operators is demonstrated via computer simulations taking into account the impact of additive noise introduced by analog devices.
\end{abstract}


\IEEEpeerreviewmaketitle

\section{Introduction}

\IEEEPARstart{A}{nalog}
signal processing has been attracting much attention
because it has a potential to realize the signal processing of large-volume data with high-speed, 
low-latency, and low-power consumption compared to conventional digital techniques\cite{b1}.
Among them, {\it optical} analog signal processing will be one of the promising approaches,
since it has been recently demonstrated that the matrix-vector product operation,
which is one of the core operations in modern signal processing,
can be efficiently realized by simply passing optical signals through optical analog
circuits \cite{b2-0}.
Thus, optical analog processing has been widely applied to problems in signal processing and machine learning \cite{b2,b3,b4,b5}.

Compressed sensing \cite{bib:comp,bib:comp2} is a mathematical framework for reconstructing an unknown sparse vector from an underdetermined linear measurement.
It has been employed in various signal processing fields,
including wireless communications and image processing.
One of the fundamental sparse signal reconstruction methods is $\ell_1$ regularization,
for which convex optimization-based algorithms using proximal splitting methods \cite{bib:prox-grad},
such as iterative shrinkage thresholding algorithm (ISTA) \cite{bib:ISTA} and fast ISTA (FISTA) \cite{bib:FISTA}, are commonly employed.
Moreover, a nonconvex regularization approach has also been explored, including 
$\ell_p$ regularization ($p \in [0,1)$) \cite{bib:l_p},
smoothly clipped absolute deviation (SCAD) regularization \cite{bib:SCAD}, 
and minimax concave penalty (MCP) regularization \cite{bib:MCP}.
Although proximal splitting methods that use these nonconvex regularizers generally do not guarantee convergence to a globally optimal solution,
they have been empirically shown to achieve superior sparse reconstruction performance compared to the $\ell_1$-regularizer.

To expand the applicable domain of compressed sensing, 
we have investigated the implementation of signal reconstruction algorithms based on ISTA, FISTA and alternating direction method of multipliers (ADMM) using optical analog circuits.
Specifically, we have developed constant inertial FISTA (CIFISTA)\cite{bib:CIFISTA}, 
a variant of FISTA with a fixed inertial parameter, 
and an approximate ADMM\cite{bib:apADMM} suited for optical analog implementation.
Moreover, we have proposed optical analog circuit architectures to realize these algorithms \cite{bib:APSIPA2023}.
In the previous works, however, we have assumed that the proximal operator of the
$\ell_1$ regularization, namely, the soft-thresholding function, is ideally implemented with some analog circuit.

\begin{figure}[tb]
\begin{center}
\includegraphics[scale=0.45]{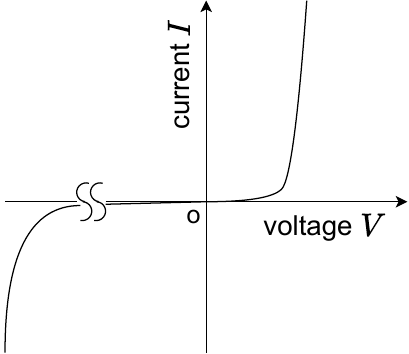}
\caption{V-I characteristics of PN-junction diode}
\label{fig:c-v}
\end{center}
\end{figure}

In this paper, with an aim to employ in optical analog compressed sensing,
we consider to realize the proximal operators of the $\ell_1$ and MCP regularization functions
with {\it electric} analog devices,
since recent OEO (optical signal - electrical signal - optical signal) 
conversion devices \cite{bib:oeo} can achieve
fast and efficient conversion of optical signals into electrical signals.
In particular, we propose to approximate the functions using PN-junction diodes \cite{bib:diode} 
inspired by the nonlinear forward voltage-current (V-I) characteristics as shown in Fig. \ref{fig:c-v}.
To demonstrate the validity of the proposed approach,
we employ the proposed approximate proximal operators for the $\ell_1$ and MCP regularization functions
in compressed sensing with ISTA, which are named {\it diode-$\ell_1$} and {\it diode-MCP}, respectively,
and evaluate their sparse reconstruction performance via computer simulations
taking into account the impact of additive noise introduced by analog devices.
The results show that both diode-$\ell_1$ and diode-MCP can achieve convergence performance comparable to
ISTA using the ideal proximal operators of the $\ell_1$ and MCP regularization functions.
Moreover, diode-$\ell_1$ achieves a steady-state mean squared error (MSE) lower than that of
ISTA using the ideal soft-thresholding function,
whereas a certain degradation of the steady-state MSE is recognized for diode-MCP compared to that
for ISTA using the ideal proximal operator of the MCP regularization function.


\section{Sparse Reconstruction with Proximal Splitting}
We consider the reconstruction problem of 
an unknown sparse vector $\bm{x}\in\mathbb{R}^N$ from its 
underdetermined linear measurement
\begin{equation}
\bm{y}=\bm{Ax}+\bm{w} \in\mathbb{R}^M, 
\label{eq:linearmodel}
\end{equation}
where $\bm{A}\in\mathbb{R}^{M\times N}$ is a known sensing matrix,
$\bm{w}\in\mathbb{R}^M$ is a measurement noise vector, and $N>M$.

A common approach for the sparse reconstruction is solving 
a regularized optimization problem of the form
\begin{equation}
\hat{\bm{x}} = \argmin_{\bm{x} \in \mathbb{R}^N} \left(  \lambda J(\bm{x}) + \frac{1}{2} \| \bm{Ax} - \bm{y} \|_2^2 \right),
\label{eq:cost}
\end{equation}
where $J(\bm{x})$ is a sparse regularizer
and $\lambda > 0$ is a regularization parameter.
Using the proximal operator of $\lambda J(\bm{x})$, the update equation of
the proximal gradient method \cite{bib:prox-grad} to solve the optimization problem
\eqref{eq:cost} is given by
\begin{align}
\bm{x}[i+1] = \text{prox}_{\epsilon \lambda J}\left(\bm{x}[i] - \epsilon \bm{A}^\mathrm{T}(\bm{Ax}[i] - \bm{y})\right),
\label{eq:proxgradmethod}
\end{align}
where $i = 0, 1, 2, \dots$ is the iteration index and
the proximal operator of a proper closed convex function 
$f: \mathbb{R}^N \to \mathbb{R} \cup \{\infty\}$ is defined as \cite{bib:proximal}
\begin{align}
\text{prox}_{\epsilon f}(\bm{v}) \coloneqq \argmin_{\bm{x} \in \text{dom}(f)}\left(  f(\bm{x}) + \frac{1}{2\epsilon} \| \bm{x} - \bm{v} \|_2^2 \right),
\end{align}
for any real number $\epsilon > 0$.
The update equation \eqref{eq:proxgradmethod} is also known as ISTA.

\subsection{$\ell_1$ Regularization}
If we choose the $\ell_1$-norm for the regularization function as $J(\bm{x})=J_{\ell_1}(\bm{x})\coloneqq\|\bm{x}\|_1$,
the problem in \eqref{eq:cost} is called $\ell_1-\ell_2$ optimization problem.
The proximal operator of the $\ell_1$-norm, 
also referred to as the soft-thresholding function, is given by
\begin{equation}
\text{prox}_{\epsilon \lambda J_{\ell_1}}(v) \coloneqq
\begin{cases}
v - \epsilon \lambda & (v > \epsilon \lambda) \\
0 & (-\epsilon \lambda \leq v \leq \epsilon \lambda) \\
v + \epsilon \lambda & (v < -\epsilon \lambda)
\end{cases}.
\label{eq:soft}
\end{equation}
When the argument of $\text{prox}_{\epsilon \lambda J_{\ell_1}}(\cdot)$ is a vector,
it is defined as a vector-valued function obtained by applying \eqref{eq:soft}
element-wise to the input vector.
In the rest of the paper, we call ISTA using the proximal operator of \eqref{eq:soft}
as {\it ISTA-$\ell_1$}.

\subsection{MCP Regularization}
In order to obtain nearly unbiased estimate, we can employ
the MCP regularization function\cite{bib:MCP} for 
$J(\bm{x})$ given by
\begin{equation}
J_\text{MCP}(x) =
\begin{cases}
|x| - \frac{x^2}{2\alpha\lambda}
&|x| \leq \alpha\lambda\\
\frac{\alpha\lambda}{2}
&|x| > \alpha\lambda
\end{cases}.
\label{eq:mcp}
\end{equation}
For $\alpha > \epsilon$,
the proximal operator of the MCP regularization function is given by
\begin{equation}
\text{prox}_{\epsilon\lambda J_\text{MCP}}(v) \coloneqq
\begin{cases}
0&(|v| < \epsilon\lambda)\\
\frac{\alpha}{\alpha - \epsilon}(v - \epsilon\lambda\operatorname{sgn}(v))
&(\epsilon\lambda \leq |v| \leq \alpha\lambda) \\ 
v&(|v| > \alpha\lambda)
\end{cases}.
\label{eq:mcp-prox}
\end{equation}
When the argument of $\text{prox}_{\epsilon\lambda J\text{MCP}}(\cdot)$ is a vector,
it is also defined as a vector-valued function obtained by applying \eqref{eq:mcp-prox} 
element-wise to the input vector.
Note that $\text{prox}_{\epsilon\lambda J\text{MCP}}(\cdot)$
remains an identity mapping for $|v| > \alpha\lambda$,
which is expected to reduce the estimation bias.
In the rest of the paper, we call ISTA using the proximal operator of \eqref{eq:mcp-prox}
as {\it ISTA-MCP}.



\section{Proposed Approximate Proximal Operators Using PN-junction Diode}
The mathematical model of the forward V-I characteristics of PN-junction diode is given by \cite{bib:diode}
\begin{align}
I_{\mathrm{D}}(V_{\mathrm{D}})&=I_{\mathrm{s}}\cdot\left(\exp\left(\frac{V_{\mathrm{D}}}{mV_{\mathrm{T}}}\right)-1\right),
\end{align}
where $I_{\mathrm{D}}$ is the current, $V_{\mathrm{D}}$ is the voltage, $I_{\mathrm{s}}$ is the saturation current, 
$m$ is the emission coefficient and $V_{\mathrm{T}}$ is the thermal voltage.

\begin{figure}[tb]
\begin{center}
\includegraphics[scale=0.8]{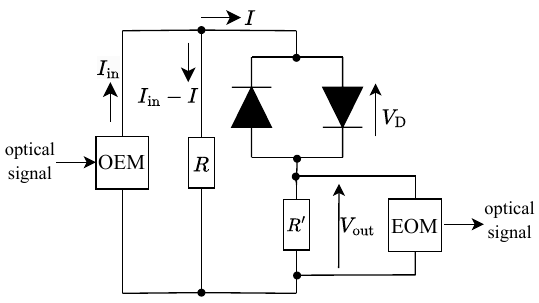}
\caption{Proposed electrical analog circuit using PN-junction diodes to approximate proximal operators.}
\label{fig:diode circuit}
\end{center}
\end{figure}

Based on the model, 
we propose an electrical analog circuit shown in Fig. \ref{fig:diode circuit}
to approximate the proximal operators.
In the figure, OEM and EOM denote the optical-electro modulator that converts optical
signals into electrical signals and the electro-optical modulator
that converts electrical signals into optical signals, respectively.
The output current of the OEM $I_\mathrm{in}$ is considered as the input to
the proposed electric analog circuit (input to the approximate proximal operator), 
while the terminal voltage $V_\mathrm{out}$ of the resistance $R'$ is the
output of the proposed circuit (output of the approximate proximal operator) 
and is also the input of the EOM.

Ignoring the current at the diodes for the reverse bias, 
the output voltage $V_{\mathrm{out}}$ as a function of $I_{\mathrm{in}}$ 
can be obtained by solving circuit equations of Fig.~\ref{fig:diode circuit} as 
\begin{align}
&V_{\mathrm{out}}(I_{\mathrm{in}})=
\mathrm{sgn}(I_{\mathrm{in}})R'
\left \{
\frac{mV_{\mathrm{T}}}{R+R'} \right .\notag \\
&~~\left . \cdot W \left( \frac{I_{\mathrm{s}}(R+R') \exp \left( \frac{I_{\mathrm{s}}(R+R') + |I_{\mathrm{in}}|R}{mV_{\mathrm{T}}} \right)}{mV_{\mathrm{T}}} \right) - I_{\mathrm{s}}
\right \},
\label{eq:vout}
\end{align}
where $W(\cdot)$ is Lambert $W$-function\cite{bib:w}.
Note that the $W$-function is a multi-valued function in general
defined in the framework of complex function theory, 
but it can be regarded as an ordinary uni-valued function by choosing its main branch
for the argument of real positive numbers, which corresponds to our case because
we ignore the reverse current at the diodes.

In the proposed function $V_{\mathrm{out}}(I_{\mathrm{in}})$ in \eqref{eq:vout}, 
only the resistances $R$ and $R'$ are freely adjustable parameters,
while $I_{\mathrm{s}}$, $m$, and $V_{\mathrm{T}}$ are constants,
which depend on the environment or the choice of diodes.
Thus, we adjust $R$ and $R'$ to approximate the proximal operators of the $\ell_1$ and
MCP regularization functions in the following subsections.

\subsection{Approximate Soft-Thresholding Function}
\label{subsec:electrical-1}
Since the ideal soft-thresholding function has a slope of 1 outside the thresholding point, 
we adjust $R$ and $R'$ so that the approximate function also has the same characteristic.
The limit values of the derivative of $V_{\mathrm{out}}$ with respect to $I_{\mathrm{in}}$ as $I_{\mathrm{in}}\rightarrow\infty$ and $I_{\mathrm{in}}\rightarrow -\infty$ are both given by
\begin{align}
\frac{dV_{\mathrm{out}}}{dI_{\mathrm{in}}}
&\rightarrow \frac{RR'}{R+R'} \quad (I_{\mathrm{in}} \rightarrow \pm \infty).
\end{align}
Thus, the characteristic can be achieved by setting $RR'/(R+R')=1$.

\begin{figure}[tb]
\begin{center}
\includegraphics[scale=0.41]{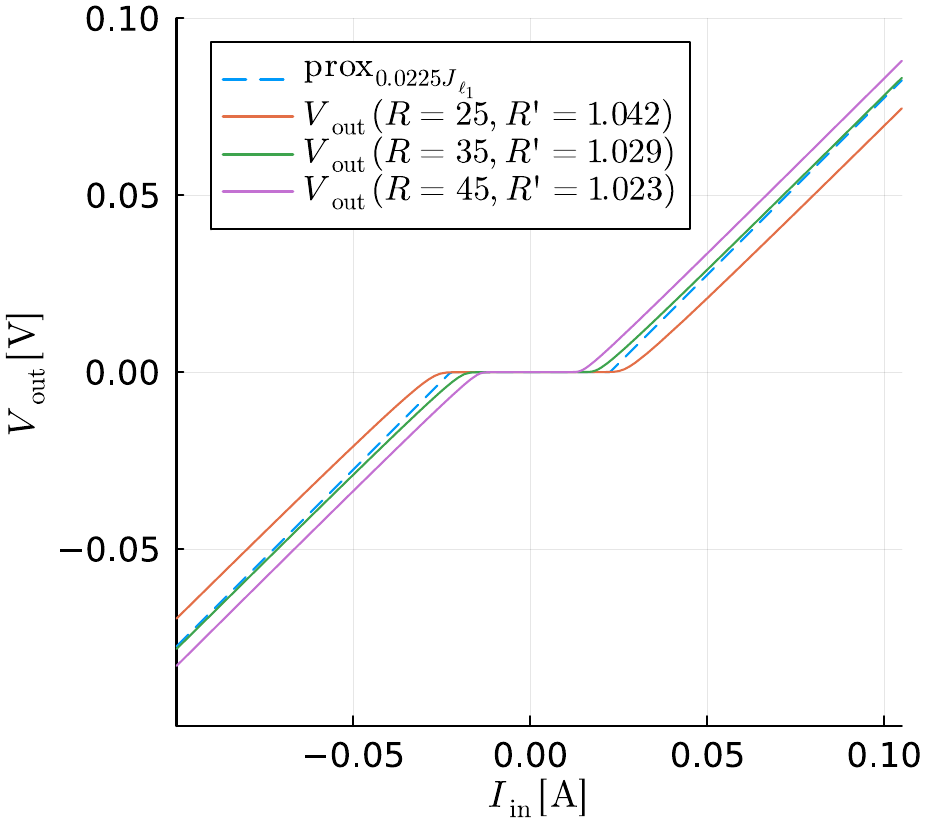}
\caption{Examples of $V_{\mathrm{out}}(I_{\mathrm{in}})$ to approximate soft-thresholding function $\text{prox}_{0.0225 J_{\ell_1}}(I_{\mathrm{in}})$.}
\label{fig:fit}
\end{center}
\end{figure}

To confirm the capability of $V_{\mathrm{out}}(I_{\mathrm{in}})$ to approximate
the soft-thresholding function,
Fig. \ref{fig:fit} shows some examples of $V_{\mathrm{out}}(I_{\mathrm{in}})$
to approximate $\text{prox}_{\epsilon \lambda J_{\ell_1}}(I_{\mathrm{in}})$ with $\epsilon\lambda=0.0225$,
where we set $I_{\mathrm{s}}=1.4 \times 10^{-14}$, $m=1$ and $V_{\mathrm{T}}=0.026$.
For the approximate soft-thresholding function,
we set $R=25,\,35,\,45\, \Omega$ and $R'$ is determined to satisfy $RR'/(R+R')=1$ for each
value of $R$, which result in $R'=1.042,\,1.029,\,1.023\, \Omega$, respectively.
From the figure, we can see that $V_{\mathrm{out}}(I_{\mathrm{in}})$ can approximate the soft-thresholding
function well with the appropriate choice of $R$ and $R'$.
Note also that those values of $R$ and $R'$ are feasible (not too large or too small) 
for practical circuit implementation.

\subsection{Approximate Proximal Operator of MCP Regularization}
\label{subsec:electrical-2}
\begin{figure}[tb]
\centering
\includegraphics[scale=0.4]{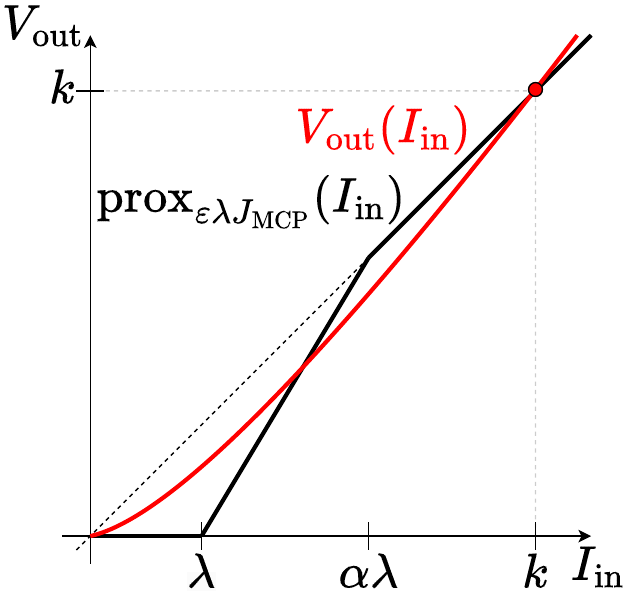}
\caption{Schematic illustration of proposed approximate proximal operator of MCP regularization function.}
\label{fig:MCP_fit}
\end{figure}

The proximal operator of the MCP regularization function is an identity map for $|I_{\mathrm{in}}| > \alpha \lambda$, therefore, it is desirable that $V_{\mathrm{out}}(I_{\mathrm{in}})$ also behaves as an identity map in the corresponding regions for the approximation.
Due to the strictly monotonic nature of Lambert $W$-function, however,
this is inherently not possible.
Therefore, we impose the condition that 
$V_{\mathrm{out}}(I_{\mathrm{in}})$ to be the identity map for two points
$I_{\mathrm{in}}=\pm k$, instead of the regions, for some $k >\alpha \lambda$ 
to reduce the bias of the estimate.
Since both $\text{prox}_{\epsilon \lambda J_\text{MCP}}(I_\text{in})$
and $V_{\mathrm{out}}(I_{\mathrm{in}})$ are odd functions,
the condition is satisfied as far as $V_{\mathrm{out}}(k)=k$ holds, 
which results in 
\begin{equation}
    R = \frac{R'}{R' - 1} \left( 1 + \frac{m V_{\mathrm{T}}}{k} \log\left(1 + \frac{k}{I_{\mathrm{s}} R'}\right) \right).
\label{eq:k}
\end{equation}  
Schematically, this approach is illustrated in Fig. \ref{fig:MCP_fit} for $I_{\mathrm{in}}>0$,
where $V_{\mathrm{out}}(I_{\mathrm{in}})$ passes through the point $(k,k)$
to approximate $\text{prox}_{\epsilon \lambda J_\text{MCP}}(I_\text{in})$.

\begin{figure}[tb]
\centering
\includegraphics[scale=0.5]{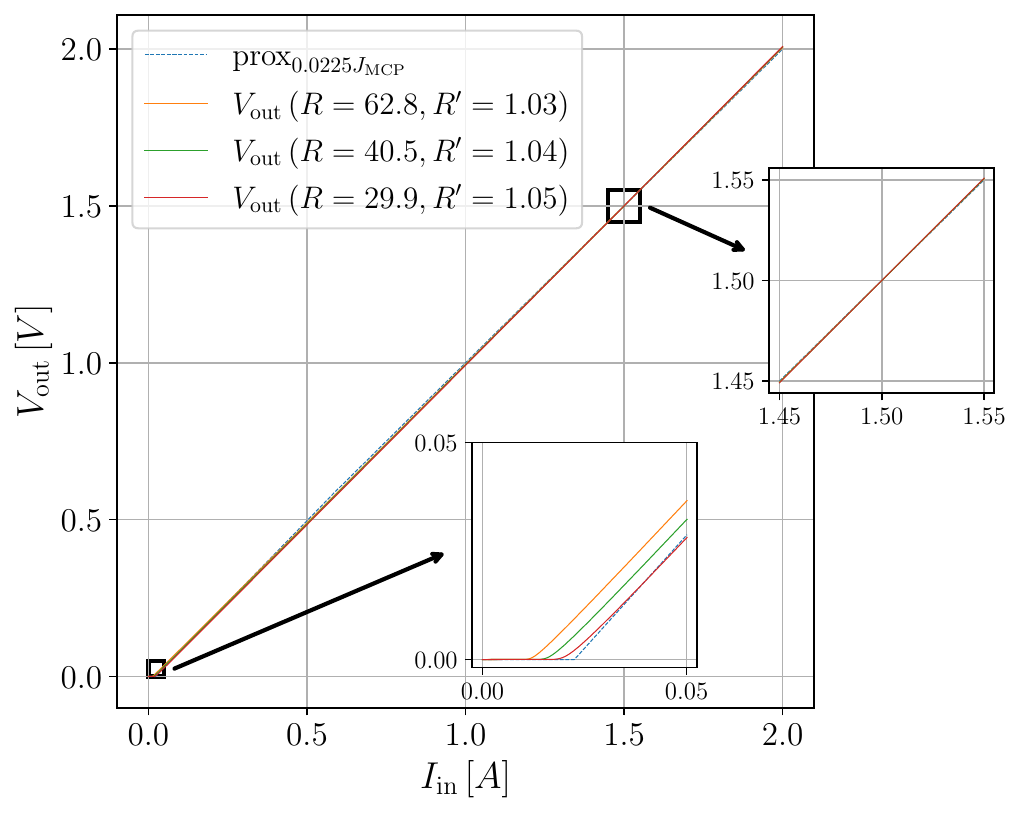}
\caption{Examples of \( V_{\mathrm{out}}(I_{\mathrm{in}}) \) to approximate proximal operator \( \text{prox}_{0.0225 J_\text{MCP}}(I_{\mathrm{in}}) \) for \( 0 \leq I_{\mathrm{in}} \leq 2 \) with  $k=1.5$.}
\label{fig:fit2}
\end{figure}

Fig.~\ref{fig:fit2}
illustrates some examples of $V_{\mathrm{out}}(I_{\mathrm{in}})$ to approximate $\text{prox}_{\epsilon \lambda J_\text{MCP}}(I_{\mathrm{in}})$
for $I_{\mathrm{in}} > 0$ with $\lambda = 0.0225$,
$\epsilon = 1$, $\alpha = 27$, $I_{\mathrm{s}} = 1.4 \times 10^{-14}$,
$m = 1$, and $V_{\mathrm{T}} = 0.026$.
In the figure, we have set $k=1.5$ and $R' = 1.03, 1.04, 1.05 \, \Omega$,
while $R$ is determined from Equation~\eqref{eq:k} as $R = 62.8, 40.5, 29.9 \, \Omega $, respectively.
Note that these values of $R$ and $R'$ are also feasible (not too large or too small) for practical implementation. 

\section{Additive Circuit Noise at OEM and EOM}
In the proposed approach, we assume the employment of 
OEM and EOM as shown in Fig.~\ref{fig:diode circuit}.
In the signal conversion process at OEM, 
thermal noise and shot noise are unavoidably superimposed, 
while thermal noise is added in the process at EOM,
which might degrade the sparse reconstruction performance.

The shot noise and the thermal noise can be modeled as 
a Poisson random variable and a zero-mean Gaussian random variable, respectively,
whose variances are respectively given by \cite{bib:noise}
\begin{align}
    \sigma_\mathrm{shot}^2&=2qI_\mathrm{in}B,\\
    \sigma_\mathrm{thermal}^2&=\frac{4k_{\mathrm{B}}TB}{R_\mathrm{load}},
\end{align}
where $q=1.602\times 10^{-19}\,\mathrm{C}$ is the elementary charge, $I_\mathrm{in}$ is the output current of the OEM, $B$ is the signal bandwidth, $k_{\mathrm{B}}=1.381\times 10^{-23}\,\mathrm{J/K}$ is Boltzmann's constant, 
$T$ is the absolute temperature and $R_\mathrm{load}$ is the resistance, which 
generates the thermal noise ($R_\mathrm{load}=R$ for the OEM and $R_\mathrm{load}=R'$ for EOM).
The signal bandwidth is given by $B=1/(2\pi CR_\mathrm{cmb})$, 
where $C$ is the capacitance of the modulator and $R_\mathrm{cmb}$ is the combined
resistance of the circuit.


\section{Numerical Results}
To verify the effectiveness of the proposed method, we numerically evaluate the signal reconstruction performance of the proximal gradient method in \eqref{eq:proxgradmethod} implemented with an optical analog circuit as in \cite{bib:APSIPA2023},
replacing the proximal operator with $V_{\mathrm{out}}(I_{\mathrm{in}})$ as
\begin{equation}
\bm{x}[i+1]=V_{\mathrm{out}}(\bm{x}[i]-\epsilon \bm{A}^\mathrm{T}(\bm{Ax}[i]-\bm{y})).
\label{eq:diode_alg}
\end{equation}
The algorithm \eqref{eq:diode_alg} is called {\it diode-$\ell_1$} if $R$ and $R'$ 
in $V_{\mathrm{out}}(I_{\mathrm{in}})$ are determined with the method in 
Sect. \ref{subsec:electrical-1}, while it is called {\it diode-MCP} if they are
chosen with the method in Sect. \ref{subsec:electrical-2}.

In the linear measurement model of \eqref{eq:linearmodel}, 
we set the system size to $(M,N)=(32,64)$, which could be feasible for optical analog implementation.
Each element of the observation matrix $\bm{A}$ independently follows a zero-mean Gaussian distribution with variance $1/M$.
Each element of the unknown sparse vector $\bm{x}$ independently follows a Bernoulli-Gaussian distribution with an occurrence probability of 0.1
and nonzero elements follow a standard Gaussian distribution.
Each element of the observation noise $\bm{w}$ independently follows a zero-mean Gaussian distribution with variance $\sigma^2 = 1.0 \times 10^{-5}$.
The regularization parameter is set to $\lambda = 0.15$, and the step size parameter for all algorithms is set to $\epsilon = 0.99 / \theta_{\text{max}}(\bm{A}^\mathrm{T}\bm{A})$,
where $\theta_{\text{max}}(\bm{A}^\mathrm{T}\bm{A})$ represents the largest eigenvalue of $\bm{A}^\mathrm{T}\bm{A}$.
The parameters in the diode model are set to $I_{\mathrm{s}} = 1.4 \times 10^{-14}$, $m = 1$, and $V_{\mathrm{T}} = 0.026$.

Based on the discussion in \cite{bib:APSIPA2023}, a zero-mean Gaussian noise with variance $3.84 \times 10^{-8}$ is added as circuit noise at the optical amplifier in each iteration for all algorithms.
The parameters for circuit noise at the OEM and EOM are set to $T = 300$ K and $B = 10$ GHz, corresponding to approximately $C \approx 159$ nF, which is feasible for implementation with existing devices of optical communications.
The initial value of the estimate for each algorithm is set to a zero vector,
and the MSE is evaluated by 1,000 simulation trials.
To achieve the best convergence performance,
we have set  $R = 35 \Omega$ and $R' = 1.029 \Omega$ for diode-$\ell_1$,
$R = 40.5 \Omega$, $R' = 1.04 \Omega$, and $k = 1.5$ for diode-MCP,
and $\alpha = 27$ for ISTA-MCP.

\begin{figure}[t]
\centering
\includegraphics[scale=0.4]{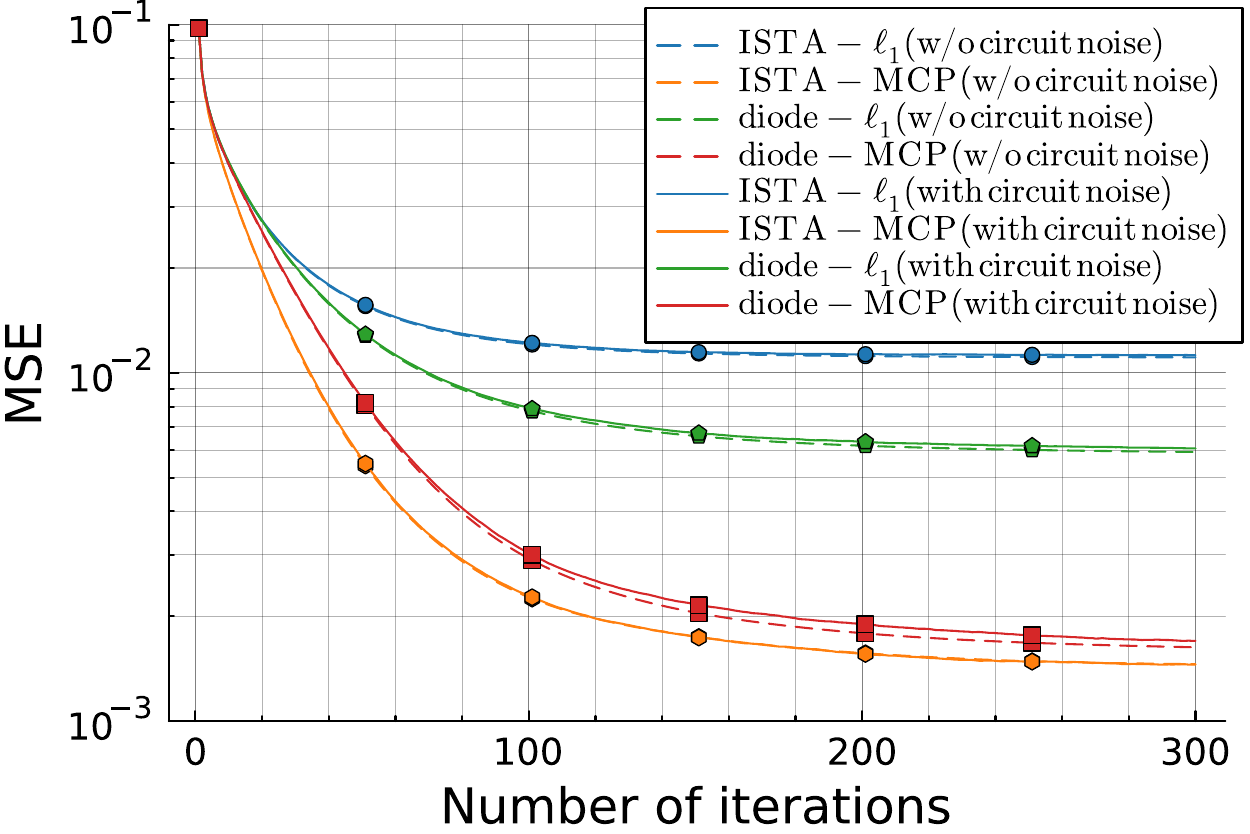}
\caption{Comparison of MSE performance (Occurrence probability of nonzero elements: 0.1)}
\label{fig:10result}
\end{figure}

Fig.~\ref{fig:10result} shows the MSE versus the number of iterations for ISTA-$\ell_1$, ISTA-MCP, diode-$\ell_1$, and diode-MCP with and without (w/o) circuit noise.
The results indicate that both diode-$\ell_1$ and diode-MCP achieve convergence rates comparable to those of 
ISTA-$\ell_1$ and ISTA-MCP, respectively.
Moreover, diode-$\ell_1$ achieves the steady-state MSE even better than ISTA-$\ell_1$.
Furthermore, diode-MCP achieves a steady-state MSE smaller than those of ISTA-$\ell_1$ and diode-$\ell_1$, 
while it is slightly worse than that of ISTA-MCP.
In addition, the impact of circuit noise on MSE performance is negligible for all algorithms,
which could indicate the feasibility of analog compressed sensing.


\section{Conclusions}
We have considered approximate functions of the proximal operators of
the $\ell_1$ norm and the MCP regularization function using forward V-I characteristics of the PN-junction diode for analog compressed sensing.
Based on the mathematical model of diodes, we have proposed
an electric analog circuit to approximate the functions, 
and demonstrated that the proposed function $V_\mathrm{out}(I_\mathrm{in})$ can approximate the proximal operators well with appropriate choices of $R$ and $R'$.
Moreover, numerical experiments of the sparse reconstruction
have shown that diode-$\ell_1$ and diode-MCP have convergence performance comparable to that of ISTA-$\ell_1$ and ISTA-MCP, respectively.
In particular, diode-MCP can achieve a steady-state MSE smaller than those of ISTA-$\ell_1$ and diode-$\ell_1$.
In addition, the impact of circuit noise on MSE performance is negligible for all algorithms.

Future work includes investigating the sensitivity of sparse reconstruction performance to the values of $R$ and $R'$, and considering the impact of other analog impairments, such as imperfect matrix-vector multiplications.


\clearpage

\clearpage

\end{document}